\documentclass[12pt]{iopart}

\usepackage{graphicx} 
\usepackage{dcolumn} 
\usepackage{bm} 

\def\sqrtsNN{\mbox{$\sqrt{s_\mathrm{NN}}$}}

\def\GeVc{\mbox{$\mathrm{GeV}/c$}}

\newcommand{ \be }{\begin{equation}}    
\newcommand{ \ee }{\end{equation}}    
\newcommand{ \bea }{\begin{eqnarray}}    
\newcommand{ \eea }{\end{eqnarray}}    
\newcommand{ \la }{\langle}    
\newcommand{ \ra }{\rangle}

\usepackage{color}

\begin{document}       

\begin{flushright}    
\end{flushright}

\vspace {-2.5cm} 
\title{Azimuthal anisotropy and correlations in p+p, d+Au and Au+Au collisions 
at 200 GeV}


\vspace {-0.2cm}

\author{A.H. Tang\dag\ for the STAR Collaboration\footnote{For the full author list and acknowledgements see Appendix "Collaborations" in this volume.}}
\address{\dag\ Physics Department, P.O. Box 5000, Brookhaven National Laboratory, 
Upton, NY~11973, aihong@bnl.gov}


\date{\today}
\vspace{-0.2cm}
\begin{abstract}
We present the first measurement of directed flow ($v_1$) at RHIC. 
$v_1$ is found to be consistent with zero at pseudorapidities $\eta$ from 
$-1.2$ to $1.2$, then rises to the level of a couple of percent over the 
range  $2.4 < |\eta| < 4$.  The latter observation is similar to data from 
NA49 if the SPS rapidities are shifted by the difference in beam rapidity 
between RHIC and SPS.  Back-to-back jets emitted out-of-plane 
are found to be suppressed more if compared to those emitted in-plane, 
which is consistent with {\it jet quenching}.
Using the scalar product method, we systematically 
compared azimuthal correlations from p+p, d+Au and Au+Au collisions.
Flow and non-flow from these three different collision systems are discussed.

\end{abstract}


\vspace{-0.2cm}

Directed flow describes the sideward bounce of the fragments 
in ultrarelativistic nuclear collisions and it carries very early 
information from the collision. Its shape at midrapidity is of special 
interest because it might reveal a signature of a possible phase transition from 
normal nuclear matter to a quark-gluon plasma~\cite{wiggle}.
Directed flow has been extensively studied at AGS and SPS, but has not previously
been reported at RHIC, because it is small and is difficult to
separate from non-flow effects. Recent RHIC results~\cite{highptsuppress}
suggest that jets are suppressed in mid-central and central 
Au+Au collisions ({\it jet quenching}). In a {\it jet quenching} picture, 
one expects that jets emitted out-of-plane suffer more energy loss than those 
emitted in-plane. The two-particle correlation function for both in- and out- 
of plane jets gives us a unique tool to test {\it jet quenching}.  Recently, 
non-flow effects in elliptic flow ($v_2$) measurements have been intensively 
discussed~\cite{nonflow}.   
A systematic comparison of azimuthal correlations among p+p, 
d+Au and Au+Au collisions will help us to
understand the non-flow contribution in Au+Au collisions. 
Probably the more important
aspect of such a study is that it can help us to understand how elliptic flow 
evolves from elementary collisions (p+p) through collisions involving cold 
nuclear matter (d+Au), and then on to hot, heavy ion collisions (Au+Au).

Two million events taken by the STAR main time projection chamber (TPC) for 
Au+Au collisions at $\sqrtsNN = 200$ GeV were used in this study,  
except for the $v_1$ analysis, which requires both the main TPC and two 
Forward Time Projection Chambers (FTPCs).  Only 70k events with FTPC data 
were available and analyzed. 

We use the three-particle cumulant method~\cite{Borghini} and event plane 
method with mixed harmonics~\cite{MarkusPoster}  in $v_1$ analysis and the results agree
with each other~\cite{MarkusPoster}. Both methods are less sensitive to 
two-particle non-flow effects because they measure three-particle correlations
\be
\la  \cos(\phi_a +\phi_b -2 \phi_{c}) \ra
= v_{1,a} v_{1,b} v_{2,c},
\label{cos3part}
\ee
in which there are no two-particle correlation terms and thus no non-flow 
contributions from them. The remaining non-flow from three-particle correlations
is expected to remain at a relative error of $20\%$, which is the major 
systematic uncertainty in this analysis. 

\begin{figure}[ht]
\begin{center}
\resizebox{
\textwidth}{!}{
\includegraphics{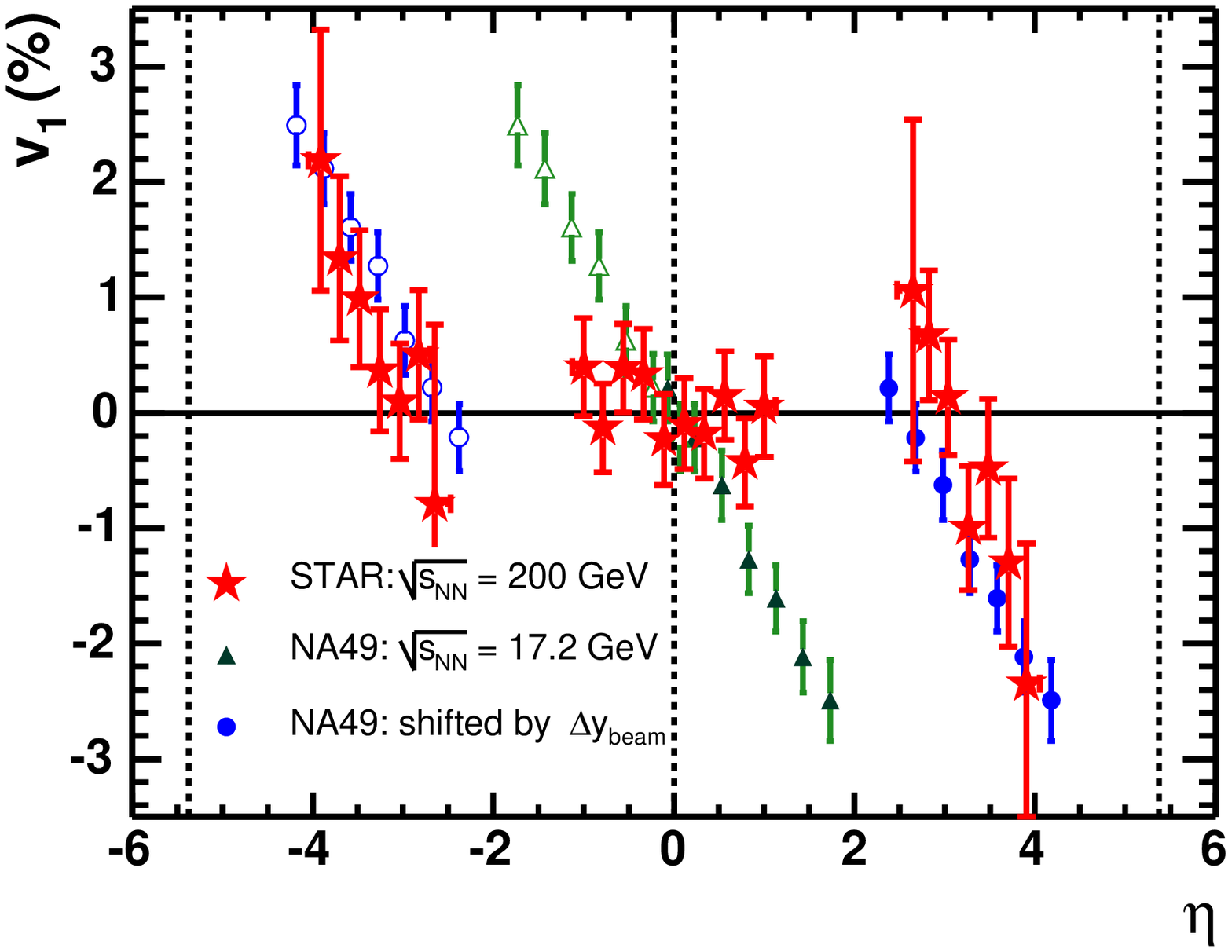}
\includegraphics{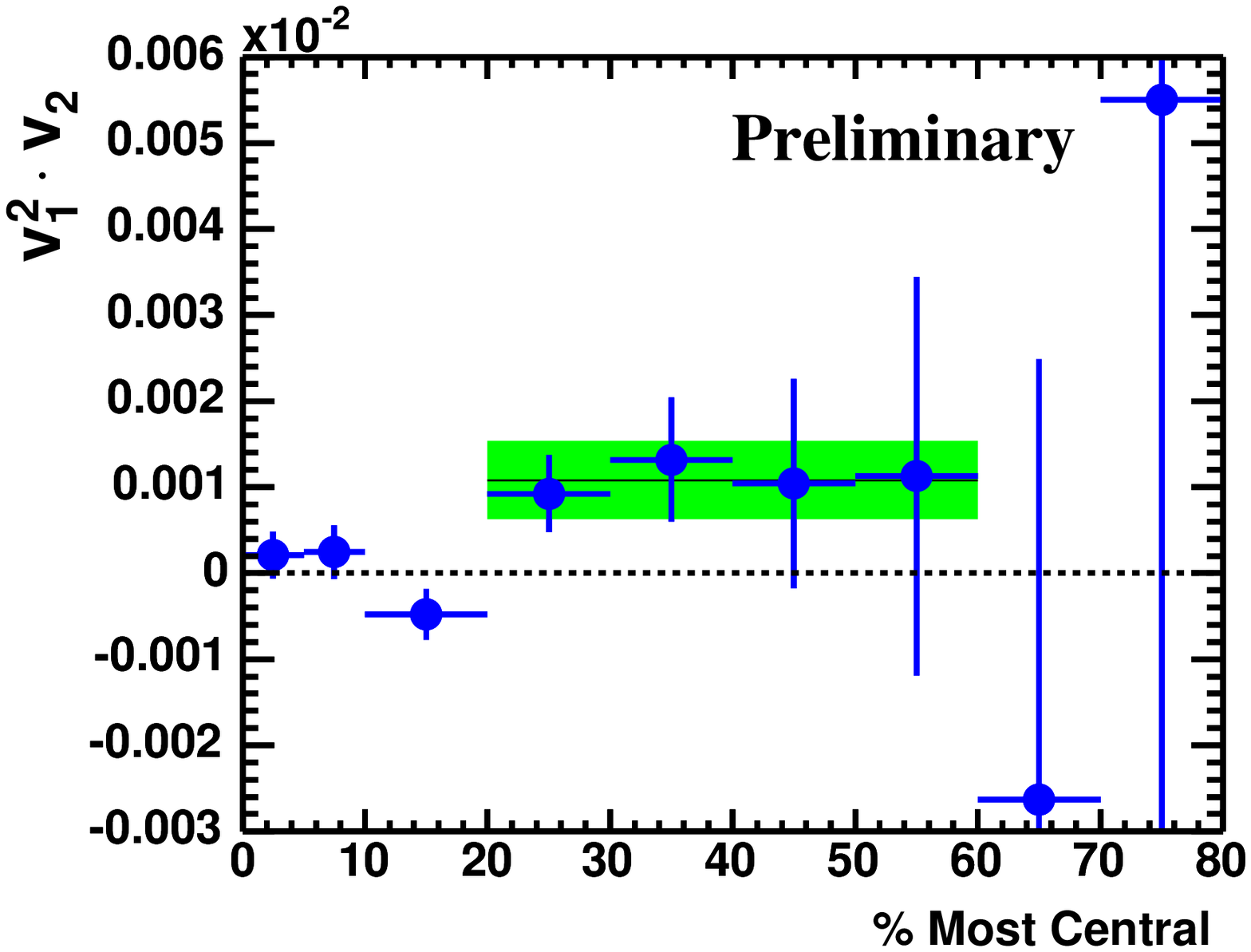}}
\vspace{-1cm}
\caption{
(left) The values of $v_1$ (stars) for charged particles for $10\%$
to $70\%$ centrality plotted as a function of pseudorapidity. Also 
shown are the results from NA49 (solid triangles) for pions from $158A$ 
GeV Pb+Pb midcentral ($12.5\%$ to $33.5\%$) collisions plotted as a 
function of rapidity. The measured NA49 points have also been shifted 
forward (solid circles) by the difference in the beam rapidities of the 
two accelerators. The open points have been reflected around midrapidity.  
The dashed lines
indicate midrapidity and RHIC beam rapidity. Both results are from analyses 
involving three-particle cumulants, $v_1\{3\}$. (right) $v_1^2v_2$ 
plotted as a function of centrality. The mean value for four mid-central 
bins is indicated by the solid line in the center of the shaded band, 
which represents the error.
\label{v1}}
\end{center}
\end{figure}

Fig.~\ref{v1} (left) shows $v_1$ from three-particle cumulants ($v_1\{3\}$) 
along with corresponding results from NA49~\cite{NA49}. 
The RHIC $v_1(\eta)$ results differ greatly from the 
directly-plotted SPS data in that they are flat near midrapidity and only
become significantly different from zero at the highest rapidities measured. 
However, when the NA49 data is replotted in terms of rapidity relative to 
beam rapidity, they look similar. In the pseudorapidity region $|\eta|<1.2$,
$v_1(\eta)$ is approximately flat with a slope of $(-0.25 \pm 0.27)$\% per 
unit of pseudorapidity, which is consistent with predictions~\cite{wiggle}. 
The quoted error is statistical only.  Note that the three-particle
correlation can also measure the sign of $v_2$ since it measures $v_1^2v_2$.
The measured correlation of Eq.~(\ref{cos3part}) is positive 
(Fig.~\ref{v1} right), and so we can conclude that we have measured the 
sign of $v_2$ to be positive. This is the 
first direct indication that the elliptic flow at RHIC is {\it in-plane}~\cite{MarkusPoster}.

\vspace{-0.5cm}

\begin{figure}
  \begin{center}
    \begin{minipage}[t]{0.48\linewidth}
\resizebox*{9.5cm}{5cm}{
\includegraphics{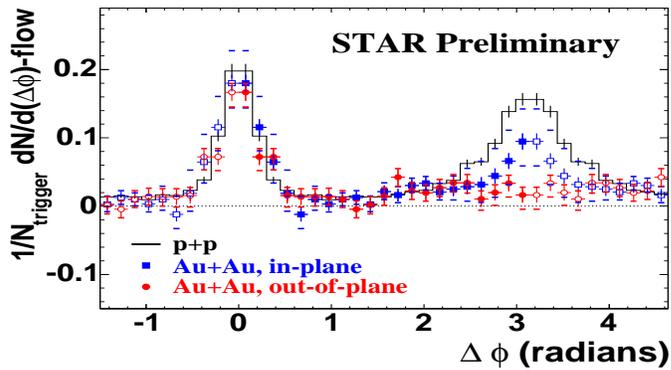}}
    \end{minipage}\hfill
    \begin{minipage}[t]{0.48\linewidth}
    \vspace{-5.5cm}  \caption{Azimuthal distributions of associated particles
 for trigger particles in-plane (squares) and out-of-plane (circles)  
for Au+Au collisions at centrality 20\%-60\%, compared with reference data 
measured in p+p collisions (histogram). The elliptic flow contribution   
is subtracted in Au+Au collisions. Open symbols are reflections of 
solid symbols  around $\Delta\phi=0$ and $\Delta\phi=\pi$.\label{fig:inOut}}
    \end{minipage}
  \end{center}
\vspace{-1cm}
\end{figure}

\vspace{-0.5cm}
Previous studies revealed large values of elliptic flow at high $p_t$~\cite{KirillHighptv2}
and strong suppression of back-to-back jets in Au+Au 
collisions~\cite{STARBackToBack}.  
Here, we extend these measurements by studying jet correlations   
with respect to the reaction plane orientation. 
We select {\it trigger} particles with  
$4<p_t^{\rm trig}<6$~GeV$/c$  
emitted in the direction of the reaction plane angle $\Psi_2$  
(in-plane, $|\phi^{\rm trig}-\Psi_2|<\pi/4$   
and $|\phi^{\rm trig}-\Psi_2|>3\pi/4$) and perpendicular to it  
(out-of-plane, $\pi/4<|\phi^{\rm trig}-\Psi_2|<3\pi/4$).  
The trigger particles are paired with {\it associated} particles with 
2 GeV$/c < p_t < p_t^{\rm trig}$. The tracks are restricted to $|\eta|<1$.
Fig.~\ref{fig:inOut} shows the azimuthal distribution of associated 
particles in Au+Au 
with elliptic flow subtracted, compared with reference data 
measured in p+p collisions. The near-side 
(around $\Delta\phi=0$) jet-like correlations measured in Au+Au are 
similar to those measured in p+p collisions. The back-to-back 
(around $\Delta\phi=\pi$) jet-like correlations measured in Au+Au 
collisions for in-plane trigger particles are suppressed compared 
to p+p, and even more suppressed for the out-of-plane trigger particles.  
The observed dependence of the suppression of the back-to-back 
correlations with respect to the reaction plane orientation is 
consistent with a jet-quenching scenario where the energy  
loss of a parton is proportional to the distance traveled through the 
dense medium. 

It is interesting to see how elliptic flow evolves from p+p collisions, 
in which non-flow dominates, through d+Au collisions, where some 
correlation with the reaction plane might develop, and finally to Au+Au 
collisions, where flow dominates. To do such a comparison,
we calculate the azimuthal correlation of particles as a function 
of $p_t$ with the entire flow vector of all particles used to 
define the reaction plane (scalar product).  
The correlation in Au+Au collisions, under the assumption that 
non-flow effects in Au+Au collisions are similar to those in p+p 
collisions, are the sum of the flow and non-flow contribution and are 
given by:  
\be   
\la u_{t} Q^{*} \ra_{AA} = M_{AA} \, v_t \,  v_Q + \la u_{t} Q^{*} \ra_{pp},  
\label{eQAA}  
\ee         
where $Q = \sum u_j$ and $Q^{*}$ its complex conjugate, $u_j 
=e^{2i\phi_j}$, $v_t$ is flow of particles with a given $p_t$, and 
$v_Q$ is the average flow of particles used to define $Q$.   
The first term in the r.h.s. of Eq.~\ref{eQAA} represents the flow   
contribution; $M_{AA}$ is the multiplicity of particles contributing 
to the $Q$ vector.  This type of variable also can be extracted from the
cumulant approach~\cite{OllitraultFlowWorkShop}.

\begin{figure}
  \begin{center}
    \begin{minipage}[t]{0.48\linewidth}
\resizebox{!}{6.5cm}{
\includegraphics{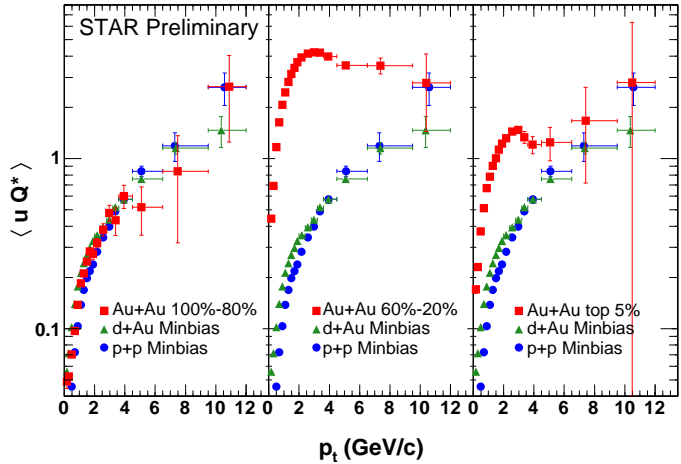}}
    \end{minipage}\hfill
    \begin{minipage}[t]{0.48\linewidth}
    \vspace{-6.5cm}  \caption{ Azimuthal correlations in Au+Au collisions 
(squares) as a function of centrality (peripheral to central in the panels 
from left to right) compared to azimuthal correlations in minimum bias p+p 
collisions (circles) and d+Au collisions (triangles). \label{fig:scalar}}
    \end{minipage}
  \end{center}
\vspace{-1cm}
\end{figure}

Fig.~\ref{fig:scalar} shows the azimuthal correlation as a 
function of transverse momentum for  
three different centrality ranges in Au+Au collisions compared to 
minimum bias d+Au collisions and minimum bias p+p collisions.  
We observe that the azimuthal correlation in peripheral Au+Au 
collisions, minimum bias d+Au collisions and minimum bias p+p 
collisions are similar to each other except at low $p_t$ ($<2$ GeV/$c$), 
where the difference is small compared to the difference 
between mid-central Au+Au collisions and the other two cases. This
is suggestive of a relatively small flow contribution in very 
peripheral Au+Au collisions and minimum bias d+Au collisions.
In mid-central events, the azimuthal correlations in Au+Au  
collisions is very different from that in p+p collisions, both in 
magnitude and $p_t$-dependence. 
For the most central Au+Au collisions, the magnitude of the correlation 
at low-$p_t$ is also different from p+p and d+Au, however, for particles 
with $p_t \ge 5$~GeV$/c$, the correlation in p+p, d+Au and Au+Au becomes the 
same within errors. This indicates that non-flow could dominate the 
azimuthal correlations in central Au+Au collisions at high $p_t$.
The centrality dependence of the azimuthal correlation in Au+Au 
collisions is clearly non-monotonic, being relatively small for very 
peripheral collisions, large for mid-central collisions, and relatively 
small again for central collisions. This non-monotonic centrality 
dependence is strong evidence that in mid-central collisions (20\%-60\%)
the measured finite $v_2$ for $p_t$ up to $7$ $\GeVc$ is due to 
real correlations with the reaction plane.

In summary, we have presented the first $v_1$ measurement at RHIC. 
$v_1(\eta)$ is found to be flat over a region on either side of midrapidity.   
We observe that back-to-back jets are suppressed more in the 
out-of-plane direction than in-plane direction, which is consistent with 
a jet quenching picture. The azimuthal correlation from p+p, d+Au and 
Au+Au collisions are compared by the scalar product method.



\end{document}